\documentclass[reprint,nofootinbib,superscriptaddress,floatfix]{revtex4-1}

\usepackage{graphicx}
\usepackage[english]{babel}
\usepackage{bbold}
\usepackage{braket}
\usepackage{graphicx}
\usepackage{float}
\usepackage{makeidx}
\usepackage{enumitem}
\usepackage{amsthm}
\usepackage{tikz}
\usepackage{color, colortbl}
\usetikzlibrary{quantikz}
\usepackage{url}
\usepackage[utf8]{inputenc}
\usepackage[caption=false]{subfig}
\usepackage{scrextend}
\usepackage[colorlinks=true,citecolor=red,linkcolor=brown,urlcolor=magenta]{hyperref}
\usepackage{makecell}
\usepackage{tabularx}
\usepackage{multirow}

\usepackage{algorithm}
\usepackage{algpseudocode}
\floatstyle{boxed}
\newfloat{algorithm}{t}{lop}
\floatname{algorithm}{Algorithm}

\DeclareMathAlphabet{\pazocal}{OMS}{zplm}{m}{n}

\newcommand{\M}{\pazocal{M}}

\newcommand{\ketbra}[2]{| #1 \rangle\langle #2 |}
\renewcommand{\t}{{\scriptscriptstyle\mathsf{T}}}

\begin{document}
\title{Quantum error mitigation by layerwise Richardson extrapolation}
\author{Vincent Russo}
\email{vincent@unitary.fund}
\affiliation{Unitary Fund}
\author{Andrea Mari}
\affiliation{Unitary Fund}
\affiliation{Physics Division, School of Science and Technology, Universit\`{a} di Camerino, 62032 Camerino, Italy}

\begin{abstract}
A widely used method for mitigating errors in noisy quantum computers is Richardson extrapolation, a technique in which the overall effect of noise on the estimation of quantum expectation values is captured by a single parameter that, after being scaled to larger values, is eventually extrapolated to the zero-noise limit. We generalize this approach by introducing \emph{layerwise Richardson extrapolation (LRE)}, an error mitigation protocol in which the noise of different individual layers (or larger chunks of the circuit) is amplified and the associated expectation values are linearly combined to estimate the zero-noise limit. The coefficients of the linear combination are analytically obtained from the theory of multivariate Lagrange interpolation. LRE leverages the flexible configurational space of layerwise unitary folding, allowing for a more nuanced mitigation of errors by treating the noise level of each layer of the quantum circuit as an independent variable. We provide numerical simulations demonstrating scenarios where LRE achieves superior performance compared to traditional (single-variable)  Richardson extrapolation.
\end{abstract}

\maketitle

\section{Introduction}

In recent years, the field of quantum technologies has witnessed extraordinary progress, especially in the evolution of noisy intermediate-scale quantum (NISQ) devices. Despite their capacity to excel over classical devices in certain tasks~\cite{arute2019quantum, zhong2020quantum, neill2018blueprint, madsen2022quantum, wu2021strong, ebadi2021quantum, daley2022practical}, NISQ devices are notably hindered by substantial noise, adversely affecting their output.

As we await the advent of fault-tolerant devices~\cite{preskill2018quantum}, a significant field of exploration for addressing the prevalent noise issues is quantum error mitigation (QEM)~\cite{cai2022quantum, li2017efficient, temme2017error, endo2018practical, kandala2019error, strikis2021learning, giurgica2020digital, larose2022mitiq, endo2021hybrid, kim2023scalable, koczor2021exponential, he2020zero, huggins2021virtual, pascuzzi2022computationally, song2019quantum}. QEM serves as an intermediate approach to fault tolerance that can be realized at present to overcome the hurdle of noisy devices. There are a variety of QEM techniques that are the subject of active research, for example, zero-noise extrapolation (ZNE)~\cite{temme2017error, li2017efficient, kandala2019error}, probabilistic-error cancellation (PEC)~\cite{temme2017error, endo2018practical, zhang2020error, van2023probabilistic}, dynamical decoupling~\cite{santos2005dynamical, viola2005random, pokharel2018demonstration, sekatski2016dynamical}, and Clifford data regression~\cite{czarnik2021error, lowe2021unified}. 

In this work, we focus on ZNE, a technique that has been used in many quantum computing experiments \cite{russo2023testing, cirstoiu2023volumetric, kandala2019error, larose2022mitiq, larose2022error, kim2023scalable} and that has shown strong performance despite the simplicity of its practical implementation. For a given quantum circuit, the primary idea of ZNE contains two steps; intentionally scaling up the noise of the circuit and then extrapolating to the noiseless limit.

For the first step, there are several techniques one can consider to intentionally increase the noise, one of which is \emph{unitary folding}~\cite{giurgica2020digital, he2020resource}; a process that increases the length of the quantum circuit, and by proxy, the noise. The second step is achieved by fitting a curve to the expectation values measured at different noise levels to extrapolate to the noiseless expectation value. One such method, \emph{Richardson extrapolation} (RE)~\cite{temme2017error}, corresponds to a single-variable polynomial interpolation of the noise-scaled expectation values.

In this work, we generalize Richardson extrapolation to a multivariate framework in which we consider multiple independent noise parameters associated with the different layers (or with different chunks) of the full circuit. We call this new approach \emph{layerwise Richardson extrapolation} (LRE), while we use the acronym RE for the conventional approach based on single-variable Richardson extrapolation. To generalize RE to the multivariate LRE technique, we need to address two sub-problems: (i) A way of scaling up the noise of specific layers, without perturbing the rest of the circuit. (ii) A way of post-processing the information obtained from the (layerwise) noise-scaled circuits to infer the zero-noise limit.

A noise-scaling strategy that can be used to solve the first sub-problem is \emph{layerwise folding}~\cite{calderon2023quantum, patel2022charter}: an approach that considers a quantum circuit as being comprised of several layers and where a variable amount of \emph{folding}~\cite{giurgica2020digital, he2020resource} can occur at any given layer of the circuit. Layerwise folding has been used in~\cite{calderon2023quantum, patel2022charter} as a circuit debugging technique, for example, to assess what layers in a quantum circuit are particularly susceptible to noise. Instead, in this work, we are not interested in using layerwise folding as an error characterization method, but as an error mitigation tool.

The second sub-problem that we need to solve is how to generalize Richardson extrapolation in a framework in which the expectation value of an observable can be considered as a multivariate function of the noise levels associated with different layers. We address this sub-problem by applying the mathematical theory of multivariate Lagrange interpolation~\cite{saniee2008simple, olver2006multivariate}, which allows us to express the zero-noise limit as a linear combination of the noise-scaled expectation values, each one weighted with a suitable real coefficient which only depends on the noise scaling factors.

It is interesting to compare the characteristic features of LRE for two similar techniques: PEC and RE. Like PEC, LRE involves a linear combination of many circuits in which only some specific layers are changed, while the rest of the circuit is kept unmodified. Unlike PEC but similar to RE, LRE does not necessitate full knowledge of the noise model. This is because the generation of modified circuits in LRE is deterministic and solely depends on the choice of the noise scale factors. It is also worth noting that for the case of linear extrapolation, LRE reduces to the noise-scaling variant of the NOX (noiseless output extrapolation) method described in the Appendix of~\cite{ferracin2022efficiently}. A further interesting connection is to the NEPEC (noise-extended probabilistic error cancellation) technique introduced in~\cite{mari2021extending}, in which noise scaling has been proposed as a way to build quasi-probability representations of individual gates (or layers) to be used for probabilistic error cancellation. Our technique is also related to~\cite{otten2019recovering}, in which ZNE has been proposed for mitigating a multi-parameter noise model. 

In~\cite{otten2019recovering}, however, the parameters are associated with different physical errors acting uniformly along the circuit (e.g. the values of $T_1$ and $T_2$ for a qubit), noise scaling is obtained by running the same circuit on different qubits, and the final extrapolation is obtained by a numerical best fit. In this work instead, we tune the noise level of different layers by using localized folding operations and without introducing additional qubits. Moreover, instead of using a numerical best fit, we provide an analytic expression for the zero-noise limit based on the theory of Richardson extrapolation.

This article is organized as follows. In Section~\ref{sec:lre}, we formally define the LRE technique and describe the noise scaling (Section~\ref{sec:lre-noise-scaling}) and extrapolation strategies (Section~\ref{sec:lre-extrapolation}). We also consider how one can apply LRE to a circuit in chunks (Section~\ref{sec:chunk-theory}) as well as the sampling overhead of LRE (Section~\ref{sec:overhead}). In Section~\ref{sec:experiments}, we showcase some examples and numerical experiments using LRE, illustrating its practical advantages and limitations. We conclude in Section~\ref{sec:discussion} with future directions and potential applications for the LRE technique.

\section{Layerwise Richardson extrapolation (LRE)}\label{sec:lre}

In this section we present the layerwise Richardson extrapolation (LRE) technique, for the mitigation of errors acting on quantum circuits. Much like RE, it consists of two major steps; noise scaling and extrapolation. For noise scaling, in Section~\ref{sec:lre-noise-scaling}, we evaluate an expectation value at different vectors of scale factors via a layerwise folding approach. For extrapolation and post-processing of these expectation values, covered in Section~\ref{sec:lre-extrapolation}, we make use of the mathematical theory of multivariate Lagrange interpolation. 

\subsection{Noise scaling}\label{sec:lre-noise-scaling}

\begin{figure}[!htpb]
    \centering
    \includegraphics[scale=1.25]{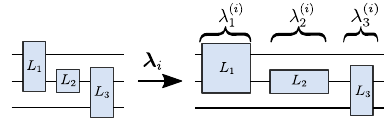}
    \caption{An arbitrary quantum circuit consisting of three layers; $L_1$, $L_2$, and $L_3$. The circuit on the right is constructed according to a vector of noise scale factors $\boldsymbol{\lambda}_i = (\lambda_1^{(i)}, \lambda_2^{(i)}, \lambda_3^{(i)})$ that determines how much the depth of each layer is scaled up by unitary folding (or by any other noise scaling method which can act layerwise). Without noise, the two circuits are equivalent. With noise, the circuit on the right is subject to more errors. Moreover, noise is amplified on some specific layers and less amplified (or unchanged) on other layers.}
    \label{fig:layerwise-folding}
\end{figure}

In RE, one of the mechanisms that is often used to scale the noise is \emph{unitary folding}~\cite{giurgica2020digital, he2020resource}. A more targeted way in which the noise can be scaled is to apply \emph{layerwise folding}, as proposed in, for instance,~\cite{calderon2023quantum, patel2022charter}. Instead of increasing the depth of the entire circuit considered as a single global entity, layerwise folding acts on specific layers of the circuit (see Figure~\ref{fig:layerwise-folding}). 

An $n$-qubit quantum circuit $C$ may be represented as a series of $\ell$ layers. Each layer $L_k$ for $1 \leq k \leq \ell$ contains one or more quantum gates acting concurrently on an $n$-qubit system
\begin{equation}\label{eq:layers}
    C = L_{\ell} L_{\ell -1} \cdots L_{2} L_{1}.
\end{equation}
In what follows, we denote each term $L_k$ as a \emph{layer}. However, the full theory of LRE is equally applicable assuming that each $L_k$ represents a multi-layer \emph{chunk} of the full circuit (see Section~\ref{sec:chunk-theory} for more details).

Consider a collection of $N$ different scale factor vectors
\begin{equation}\label{eq:scale-factor-vectors}
    \Lambda = \{
        \boldsymbol{\lambda}_1, 
        \boldsymbol{\lambda}_2,
        \ldots,
        \boldsymbol{\lambda}_N
    \},
\end{equation}
where each $\boldsymbol{\lambda}_i$ is a vector of $\ell$ scale factors that  specifies how the noise is scaled across different layers
\begin{equation}
    \boldsymbol{\lambda}_i = \left(\lambda_1^{(i)}, \lambda_2^{(i)}, \ldots, \lambda_{\ell}^{(i)}\right), \quad \lambda_k^{(i)} \ge 1.
\end{equation}
For a collection of scale factor vectors defined by $\Lambda$, we denote
\begin{equation}\label{eq:circuit-vector}
    C^{\Lambda} = 
    \{ 
        C^{\boldsymbol{\lambda}_1}, 
        C^{\boldsymbol{\lambda}_2},
        \ldots,
        C^{\boldsymbol{\lambda}_N}
    \}
\end{equation}
as the corresponding set of circuits. Each circuit in $C^{\Lambda}$ is layerwise noise-scaled according to the corresponding scale factor vector.

While layerwise folding is our chosen method for scaling the noise, it is important to note that this approach is not the only option. In principle, various methods can be employed to selectively scale the noise of specific circuit layers. For example, a promising alternative is given by the pulse-stretching method \cite{temme2017error, kandala2019error}, assuming the possibility of applying different stretchings to different layers.

For layerwise folding, each scale factor $\lambda_k^{(i)}$ corresponds to the $k$-th layer $L_k$ of the circuit $C$ and is defined as
\begin{equation}\label{eq:m_to_lambda}
    \lambda_k^{(i)} = 1 + 2 m_k^{(i)}.
\end{equation}
Here, $m_k^{(i)}$ is a non-negative integer representing the number of times the $k$-th layer $L_k$ is to be folded. The folding operation~\cite{giurgica2020digital, he2020resource} for each layer $L_k$ is expressed as
\begin{equation} \label{eq:global_folding}
L_k^{\lambda_k^{(i)}} = \left(L_k L_k^\dagger\right)^{m_k^{(i)}} L_k,
\end{equation}
where $L_k^{\lambda_k^{(i)}}$ is the new $k$-th layer after the folding operation. If $L_k$ represents a chunk of the circuit that is itself composed of $t$ elementary sub-layers $L_k= G_{k,t} \cdots G_{k,2} G_{k,1}$, unitary folding can be applied in different ways. One option,  known as \emph{global folding}~\cite{giurgica2020digital}, corresponds to Equation~\eqref{eq:global_folding}. A common alternative option, known as \emph{local folding}~\cite{giurgica2020digital}, is instead:
\begin{equation} \label{eq:local_folding}
L_k^{\lambda_k^{(i)}} = (G_{k,t}G_{k,t}^\dag)^{m_k^{(i)}}G_{k,t} \cdots (G_{k,1}G_{k,1}^\dag)^{m_k^{(i)}}G_{k,1}.
\end{equation}
Both methods scale the depth of $L_k$ by $\lambda_k^{(i)}\ge 1$ and are exactly equivalent when $t=1$. For large $t$, one can apply unitary folding partially or randomly \cite{giurgica2020digital, he2020resource}, such that the scale factor $\lambda_k^{(i)}$ is not constrained to take the odd integer values as implied by Equation~\eqref{eq:m_to_lambda}. However, for simplicity, in this work, we always assume odd integer scale factors since they are always implementable for any $t$. In this sense, each sub-layer $G_{k,j}$ in the local folding scheme can be treated as an elementary unit, analogous to how the entire layer $L_k$ is treated in the global folding scheme.

For a given vector  $\boldsymbol{\lambda}_i$ of scale factors,\ the resulting circuit $C^{\boldsymbol{\lambda}_i}$ is represented as
\begin{equation} \label{eq:layerwise-folding}
    C^{\boldsymbol{\lambda}_i} = L_\ell^{\lambda_\ell^{(i)}} 
    L_{\ell - 1}^{\lambda_{\ell -1}^{(i)}}
    \cdots 
    L_{2}^{\lambda_{2}^{(i)}} 
    L_{1}^{\lambda_{1}^{(i)}}.  
\end{equation}
In this framework, the vector of scale factors 
$\boldsymbol{\lambda}_i$ explicitly defines which layers of the circuit are to be folded and the number of times each specified layer is folded. A depiction of the folding operation for an arbitrary circuit is shown in Figure~\ref{fig:layerwise-folding}.

For each circuit in $C^{\Lambda}$ from Equation~\eqref{eq:circuit-vector}, one may compute the corresponding expectation value of a fixed observable of interest $O$. Specifically, we denote all the  expectation values associated with  $C^{\Lambda}$ as
\begin{equation}\label{eq:exp-value-vector}
    \mathbf{z} = 
    \left(
        \langle O(\boldsymbol{\lambda}_1) \rangle,
        \langle O(\boldsymbol{\lambda}_2) \rangle,        
        \ldots,
        \langle O(\boldsymbol{\lambda}_N) \rangle
    \right)^{\t}
\end{equation}
where $\langle O(\boldsymbol{\lambda}_i) \rangle$ is the expectation value of the observable $O$, estimated from the execution of the circuit $C^{\boldsymbol{\lambda}_i}$.

\begin{figure*}[!htpb]
      \centering
    \includegraphics[width=1.0\textwidth]{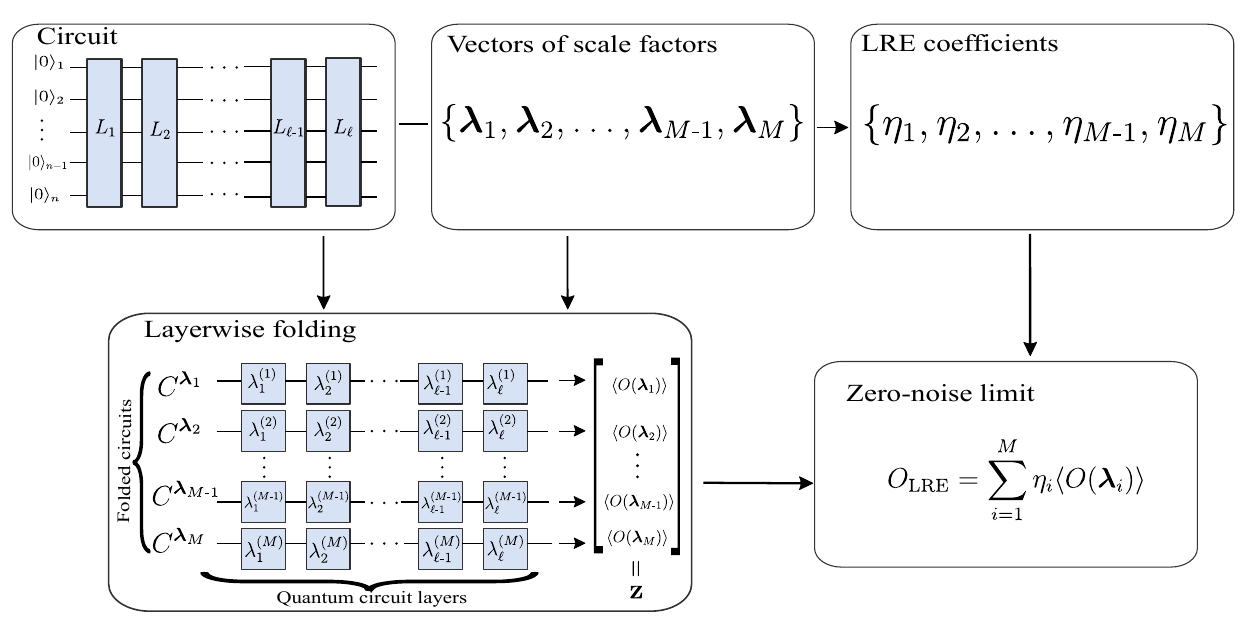}
     \caption{\small An overview of the LRE experimental workflow. As input, we consider an $n$-qubit quantum circuit consisting of $\ell$ layers or, equivalently, $l$ circuit chunks. Given the parameter $l$ and the extrapolation order $d$, we generate $M = \binom{d + l}{d}$ linearly-independent vectors of scale factors (see Equation~\eqref{eq:lambda_pattern} for a convenient generation pattern). In practice, for each vector of scale factors, one can set most elements to $1$ (no noise scaling) and assign larger values to just a few elements. From this, we perform layerwise folding on the input circuit generating $M$ different circuits, one for each vector of scale factors. Each generated circuit is almost identical to the input one, except for a few layers that are folded to amplify their noise sensitivity. For each resulting circuit (Equation~\eqref{eq:circuit-vector}), we experimentally estimate the respective expectation value (Equation~\eqref{eq:exp-value-vector}). The linear combination coefficients $\{ \eta_j \}$ can be computed straightforwardly from the multivariate Lagrange interpolation formula (Equation~\eqref{eq:eta-coefficients}) and, remarkably, they only depend on the scale factor vectors. By taking a linear combination of the noise-scaled expectation values, we obtain the error-mitigated result.}
     \label{fig:LRE}
\end{figure*}

\subsection{Extrapolation}\label{sec:lre-extrapolation}

Once we have scaled the noise via the layerwise folding approach discussed in Section~\ref{sec:lre-noise-scaling} and obtained a vector of expectation values evaluated at different vectors of scale factors as in Equation~\eqref{eq:exp-value-vector}, we proceed to post-process this raw data by a multivariate generalization of Richardson extrapolation. 

For a vector of scale factors $\boldsymbol{\lambda} = (\lambda_1, \ldots, \lambda_{\ell})$, we define the basis of all monomial terms of $\ell$ variables  of maximum degree $d$ as $\M(\boldsymbol{\lambda}, d)$. 
For instance, for $\boldsymbol{\lambda} = (\lambda_1, \lambda_2)$ and $d=2$, we have 
\begin{equation}
    \M(\boldsymbol{\lambda}, 2) = 
    \{
       1, \lambda_1, \lambda_2, \lambda_1^2, \lambda_1 \lambda_2, \lambda_2^2
    \}.
\end{equation}
In general, the number of monomial terms is given by
\begin{equation}\label{eq:number-of-monomials}
    M \equiv \left|\M(\boldsymbol{\lambda}, d)\right| = 
    \binom{d + \ell}{d},
\end{equation}
and we assume that the monomials are ordered with an increasing total degree. For example, a typical choice is the graded lexicographic order~\cite{cox2013ideals}. This implies that the first element of the list of monomials is 1, i.e., the term of zero degree that survives when taking the zero-noise limit $\boldsymbol{\lambda} \rightarrow \boldsymbol{0}$, where $\boldsymbol{0}$ is the all-zero vector.

For our purposes, typical values of the maximum degree are $d = 1$ or $d = 2$, corresponding to a linear scaling $M = \ell + 1$ and a quadratic scaling $M = (\ell + 1)(\ell + 2)/2$ of the number of terms, respectively. More generally, for a fixed extrapolation order $d$, the number of monomials $M$ scales polynomially with respect to $\ell$ since we have
\begin{equation}
    M = \frac{1}{d!} \prod_{i=1}^d \left(\ell + i\right) = \mathcal{O}(\ell^d).
\end{equation}

We aim to interpolate a multivariate polynomial function that captures the relationship between the vectors of scale factors and the expectation values as defined from Equation~\eqref{eq:exp-value-vector}. Specifically, we model the dependence of the expectation value as a function of the noise scale factors as a generic polynomial of degree $d$
\begin{equation}\label{eq:multivariate-polynomial-function}
    \langle O(\boldsymbol{\lambda}) \rangle = \sum_{j=1}^M c_j \M_j(\boldsymbol{\lambda}, d),
\end{equation}
where $\{c_j\}$ are real coefficients. We are particularly interested in extrapolating Equation~\eqref{eq:multivariate-polynomial-function} to the zero-noise limit, that is
\begin{equation}
    O_{\text{LRE}} \equiv 
    \langle O({\mathbf{0}}) \rangle = \sum_{j=1}^M c_j \M_j(\boldsymbol{0}, d) = c_1.
\end{equation}

Given the collection $\Lambda$ of scale factor vectors, as defined in Equation~\eqref{eq:scale-factor-vectors}, we define the \emph{sample matrix} 
\begin{equation}\label{eq:sample-matrix}
    \mathbf{A}(\Lambda, d) = 
    \begin{bmatrix}
        a_{1,1} & a_{1,2} & \cdots & a_{1,M} \\
        a_{2,1} & a_{2,2} & \cdots & a_{2,M} \\
        \vdots & \vdots & \ddots & \vdots \\
        a_{N,1} & a_{N,2} & \cdots & a_{N,M}
    \end{bmatrix},
\end{equation}
where each entry in the matrix is defined as
\begin{equation}
    a_{i,j} = \M_j(\boldsymbol{\lambda}_i, d).
\end{equation}
As a notational convention, we often write Equation~\eqref{eq:sample-matrix} as just $\mathbf{A}$, whenever it is clear what the values of $\Lambda$ and $d$ are. Each row of $\mathbf{A}$ corresponds to a specific scale factor vector,  while each column corresponds to a specific monomial. The interpolation problem can be cast as a linear system, 
\begin{equation}
    \mathbf{A} \mathbf{c} = \mathbf{z},
\end{equation}
where $\mathbf{z}$ is the known vector of noise-scaled expectation values as defined in Equation~\eqref{eq:exp-value-vector} and $\mathbf{c} = (c_1, \ldots, c_M)^{\t}$ is the unknown vector of coefficients defined in Equation~\eqref{eq:multivariate-polynomial-function}. In principle, solving for $\mathbf{c}$, one can determine all the coefficients of the interpolating polynomial, which can be used to evaluate new domain points, including the zero-noise limit ($\langle O({\mathbf{0}}) \rangle = c_1$). However, if we are only interested in the zero-noise limit, it is not necessary to evaluate the full vector of coefficients $\mathbf{c}$. We will use the theory of Lagrange interpolation to obtain a simple formula that directly provides the zero-noise limit.

To have a unique solution for the system of equations, we assume that the sample matrix is square and that its determinant is non-zero. This implies that the number $N$ of different scale factor vectors is not arbitrary but it must be equal to the number of monomials, i.e.,
\begin{equation}
 N = M \quad \text{and} \quad  \det\left(\mathbf{A}(\Lambda, d)\right) \not= 0.
\end{equation}
In practice, for a given number of layers $\ell$ and a given degree $d$ of the interpolating polynomial, the number of different noise scaling configurations and the number of different expectation values that one needs to measure is given by Equation~\eqref{eq:number-of-monomials}. Note that assuming $\det(\mathbf{A}) \not= 0$ is not a strong limitation since, in the case of a zero (or close to zero) determinant, one can always change some of the scale factor vectors in such a way to avoid an ill-conditioned system of equations.

By a direct application of the theory of multivariate Lagrange interpolation (as shown in Appendix~\ref{sec:multivariate-interpolation}), we can obtain 
 the zero-noise limit via the following linear combination of the noisy expectation values
 \begin{equation}\label{eq:lre-linear-combo}
    O_{\text{LRE}} = 
    \sum_{i=1}^{M} \eta_i \langle O(\boldsymbol{\lambda}_i) \rangle,
\end{equation}
where the coefficients are given by
\begin{equation}\label{eq:eta-coefficients}
    \eta_i = 
\frac{\det \left(\mathbf{M}_i\right)}{\det \left(\mathbf{A}\right)}, 
\end{equation}
where $\mathbf{M}_i$ is the matrix obtained from $\mathbf{A}$ after replacing the $i$-th row by the vector $\mathbf{e}_1 = (1, 0, \ldots, 0)$ consisting of a 1 followed by zeros.

\subsection{Applying LRE to chunks of the circuit}\label{sec:chunk-theory}

As we anticipated in Section~\ref{sec:lre-noise-scaling}, if instead of decomposing the circuit into a sequence of elementary layers (of depth one) we split the circuit into chunks of arbitrary depth, the whole theory of LRE is equally applicable. Indeed, in the theoretical derivation developed in the previous sections, we never had to invoke any assumption on the actual depth of each term $L_k$ in Equation~\eqref{eq:layers}.

In practice, this means that the total number of chunks $l$ in Equation~\eqref{eq:layers} is an arbitrary hyperparameter of LRE that we are free to choose at our convenience. This flexibility allows us to interpolate from $l=1$ corresponding to traditional (single-chunk) RE, up to $l=l_{\rm max}$, where $l_{\rm max}$ is the maximum number of elementary layers of the circuit. 

Operationally, given a circuit $C$ of depth $l_{\rm max}$ and a target observable $O$, the implementation of LRE corresponds to the following protocol (see also Figure~\ref{fig:LRE}):
\begin{enumerate}
    \item Choose the hyperparameters: the extrapolation order $d$, the number of splittings $l \le l_{\rm max}$, and the minimum noise scaling gap $\Delta$. By default, use $\Delta=2$ (minimum gap allowed by unitary folding). See   Section~\ref{sec:overhead} for more details on hyperparameters.
    
    \item Compute the number $M$ of degrees of freedom using Equation~\eqref{eq:number-of-monomials}.
    \item Choose $M$ different vectors of scale factors $\boldsymbol{\lambda}_1, \boldsymbol{\lambda}_2, \ldots, \boldsymbol{\lambda}_M$. A simple choice is the following
    \begin{equation}\label{eq:lambda_pattern}
    \boldsymbol{\lambda}_i = \boldsymbol{1} +  \boldsymbol{m}_i  \Delta, \quad i=1, 2, \ldots, M,
    \end{equation}
    where $\boldsymbol{1}=(1, 1, \dots)$ and $\{\boldsymbol{m}_i\}$ are all the vectors of $l$ non-negative integers with $\|\boldsymbol{m}_i\|_1 \le d$.
    
    \item Evaluate the corresponding $M$ real coefficients $\eta_1, \eta_2, \dots, \eta_M$ using Equation~\eqref{eq:eta-coefficients}.
    
    \item Split C into $l$ chunks and apply layerwise folding as defined in Equations~(\ref{eq:m_to_lambda}-\ref{eq:layerwise-folding}), generating $M$ noise-scaled circuits $C^{\boldsymbol{\lambda}_1}, 
            C^{\boldsymbol{\lambda}_2},
            \ldots,
            C^{\boldsymbol{\lambda}_M}$.
    \item Evaluate the corresponding $M$ expectation values $\langle O(\boldsymbol{\lambda}_1) \rangle,
            \langle O(\boldsymbol{\lambda}_2) \rangle,        
            \ldots,
            \langle O(\boldsymbol{\lambda}_M) \rangle$ on the quantum computer.
    
    \item Compute the error-mitigated result using $O_{\rm LRE}=\sum_{i=1}^{M} \eta_i \langle O(\boldsymbol{\lambda}_i) \rangle$.
    \end{enumerate}

Note that only Step 6 of the above protocol involves the actual usage of a quantum computer, all the other steps are just classical pre- or post-processing.

\subsection{Sampling overhead of LRE}\label{sec:overhead}

The error-mitigated expectation value obtained from layerwise Richardson extrapolation is subject to statistical uncertainty. Each noisy expectation value in the right-hand side of Equation~\eqref{eq:lre-linear-combo} must be measured with a finite number of shots and, therefore, each term will be subject to a statistical error (shot noise). After taking the linear combination, the left-hand side of the equation will be subject to statistical uncertainty due to the propagation of the statistical error of each term on the right-hand side.

For a fixed target of statistical error, the total number of shots $s_{\rm tot}$ required to evaluate Equation~\eqref{eq:lre-linear-combo} is larger than the number of shots $s_{\rm u}$ required to directly estimate the unmitigated expectation value $\langle O(\boldsymbol{\lambda})
\rangle|_{\boldsymbol{\lambda}=\boldsymbol{1}}$.
The sampling overhead required to apply LRE is captured by the ratio $s_{\rm tot}/s_{\rm u}$. Assuming all the noisy expectation values of Equation~\eqref{eq:lre-linear-combo} have equal variance and that they are measured with the same number of shots $s_{\rm tot}/M$, it is easy to show~\cite{cai2022quantum} that:
\begin{equation}\label{eq:cost_two_norm}
    \tilde{c} :=\frac{s_{\rm tot}}{s_{\rm u}} = M \tilde{\gamma}^2, \quad \tilde{\gamma} := \left (\sum_{i=1}^M |\eta_i|^2\right)^{\frac{1}{2}}, \, s_i = \frac{s_{\rm tot}}{M}.
\end{equation}

However, the sampling overhead can be reduced by using more shots on the terms that are more ``important" in the linear combination of Equation~\eqref{eq:lre-linear-combo}. For a fixed total budget of shots $s_{\rm tot}$, it is more convenient to invest $s_i \propto |\eta_j|$ shots when estimating each noise-scaled expectation value $\langle O(\boldsymbol{\lambda}_i) \rangle$. In this case, we have \cite{cai2022quantum, temme2017error}:
\begin{equation}\label{eq:cost_one_norm}
    c:=\frac{s_{\rm tot}}{s_{\rm u}}= \gamma^2, \quad \gamma :=\sum_{i=1}^M |\eta_i|, \quad  s_i = \frac{s_{\rm tot} |\eta_i|}{\gamma}.
\end{equation}

\begin{figure}[!htpb]
    \centering
    \includegraphics[scale=0.5]{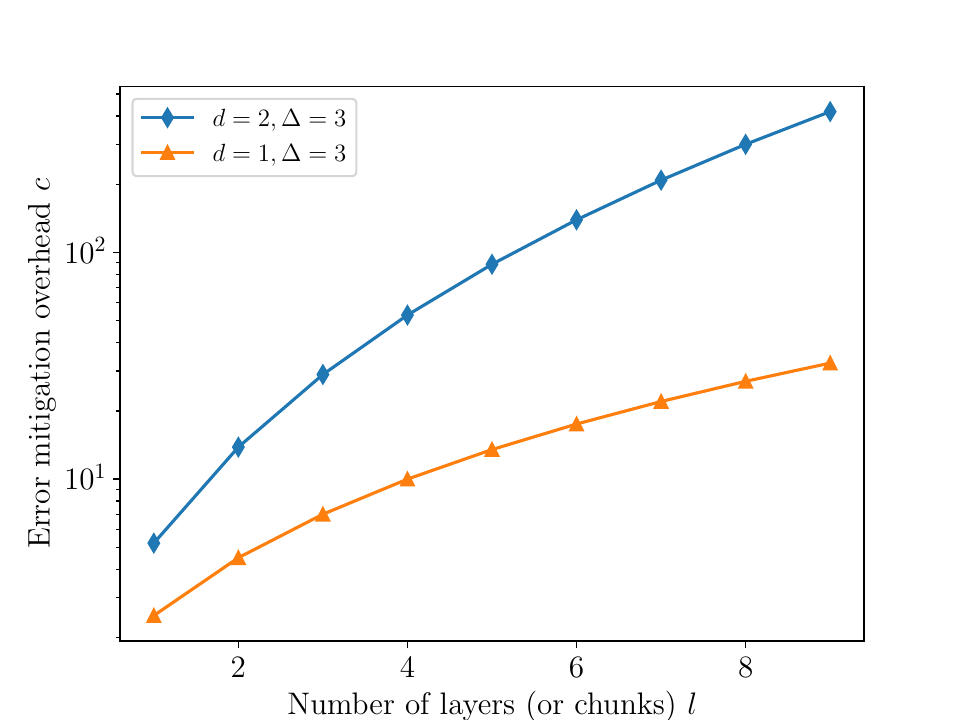}
    \caption{Sampling overhead of layerwise Richardson extrapolation for quadratic ($d=2$) and linear ($d=1$) interpolation as a function of the number of layers (or circuit chunks). The overhead is estimated according to Equation~\eqref{eq:cost_one_norm} assuming the specific choice of scale factors given in Equation~\eqref{eq:lambda_pattern}, with $\Delta=2$ (the minimum gap achievable via layerwise folding). The noise of each circuit chunk is scaled by \emph{local folding} as defined in Equation~\eqref{eq:local_folding}.}
    \label{fig:overhead}
\end{figure}

The fact that the one norm $\gamma$ of the linear combination of coefficients in \eqref{eq:lre-linear-combo} is related to the error mitigation overhead is well-known in the error mitigation literature \cite{cai2022quantum, temme2017error, takagi2022fundamental}, and it is a consequence of Hoeffding's inequality applied within the context of probabilistic  Monte-Carlo algorithms. Here we have just confirmed that the same result also holds for deterministic LRE, assuming that each expectation value in \eqref{eq:lre-linear-combo} is measured with the appropriate number of shots. 
As a direct consequence of the Cauchy-Schwartz inequality (see e.g.~\cite{cai2022quantum}), we have $\tilde{c} > c$, meaning that Equation~\eqref{eq:cost_one_norm} is the appropriate figure of merit for the optimal sampling cost.
On the other hand, in real experiments, it can be more practical to estimate noisy expectation values with the same number of shots $s_{\rm tot}/M$ for each noise-scaled circuit. In this case, $\tilde{c}$ is a more appropriate estimate of the sampling cost.

In Figure~\ref{fig:overhead} we plot $c$ as a function of the number of layers (or circuit chunks)  $l$ and for different values of the extrapolation order $d$. In this figure, we keep fixed the minimum gap between scale factors $\Delta=2$, corresponding to the minimum gap of noise scaling achievable with folding operations.

\subsubsection{Methods for reducing the sampling overhead}

\begin{figure}[!htpb]
    \centering
    \includegraphics[scale=0.5]{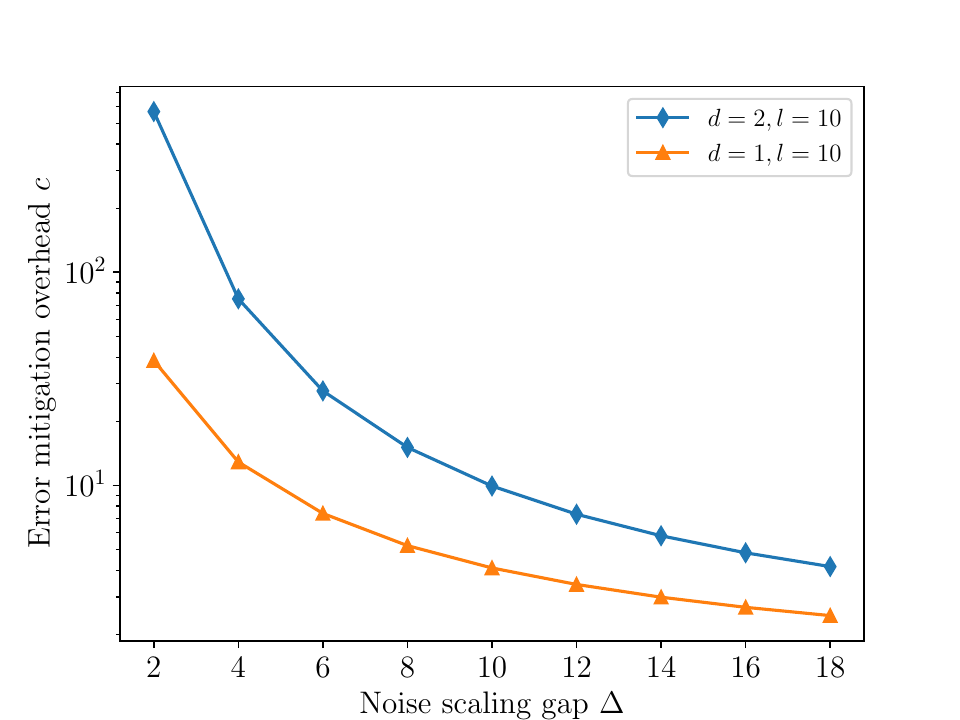}
    \caption{Sampling overhead of layerwise Richardson extrapolation for quadratic ($d=2$) and linear ($d=1$) interpolation as a function of the minimum gap between scale factors  $\Delta$. For both curves, we assume the same number of layers (or circuit chunks) $l=10$. The vectors of scale factors are chosen according to Equation~\eqref{eq:lambda_pattern}.}
    \label{fig:overhead_vs_lambda_min}
\end{figure}

In Figure~\ref{fig:overhead_vs_lambda_min}, we fix $l=10$ and show the dependence of the sampling overhead as a function of the minimum gap between scale factors $\Delta = 2, 4, 6, \dots $ corresponding to a gap in the number of folding operations equal to $1, 2, 3, \dots$, respectively (see Equation~\eqref{eq:m_to_lambda}). We observe that using a large gap between scale factors reduces the sampling cost. On the other hand, high values of noise scaling can increase the bias of the polynomial extrapolation, since the noisy expectation value is sampled further away from the zero-noise limit. Therefore, by altering $\Delta$ one can change the variance-bias tradeoff of the error-mitigated result.

Another simple way of reducing the overhead is by splitting the full circuit into a smaller number of chunks $l$, where each chunk contains multiple elementary layers. From Figure~\ref{fig:overhead}, it is clear that using a small value of $l$ is a direct way of reducing the sampling cost. 

In practice, even for very deep circuits, we can always keep the overhead of LRE under control by setting an upper bound to the number of splittings $l$ or by increasing $\Delta$, at the cost of increasing the estimation bias (see Sections~\ref{sec:ghz-circuits-scale-factor-multiplier} and \ref{sec:chunk-experiments} for numerical examples).

\section{Numerical experiments}\label{sec:experiments}
In the previous section, we presented the theory of layerwise Richardson extrapolation. In this section, we test the technique with several numerical experiments to understand its practical advantages and its limitations. In particular, we focus on a systematic comparison between LRE and traditional single-variable Richardson extrapolation (RE). 

A convenient choice of circuits for benchmarking error mitigation strategies are those which, without noise, restore all the qubits to the initial state $\ket{00\dots}$. In this case, by taking as a target observable the projector on the zero state, i.e. $O = \ketbra{00\dots}{00\dots}$, the ideal expectation value is always equal to 1 for a noiseless quantum computer. For a noisy backend instead, we can quantify the performance of different mitigation strategies by checking how close their associated predictions are to the ideal value of 1. For all of the quantum circuits simulated in this section, we assume a local amplitude damping noise model as described in Appendix~\ref{sec:noise-model}. 

In our analysis (which is depicted in Figures~\ref{fig:ghz-num-qubits}, \ref{fig:ghz-num-degree}, \ref{fig:ghz-num-shots},  \ref{fig:ghz-scale-factor-multiplier}, \ref{fig:ghz-num-chunks}, and~\ref{fig:LRE-random-circuit}), we always fix the same total budget of shots $s_{\rm tot}$ that must be used by each error mitigation strategy (trivial unmitigated, LRE, RE, etc.). This means that if an error mitigation technique requires running $M$ circuits, the total budget of shots is optimally split among the $M$ circuits such that the total sum of circuit executions is kept constant, i.e. $\sum_{i=1}^M s_i = s_{\rm tot}$. For both LRE and RE, we use the optimal splitting $s_i$ defined in Equation~\eqref{eq:cost_one_norm} (recalling that RE is a special case of LRE with $l=1$).  If not explicitly specified, we fix a total budget of $s_{\rm tot}=10^6$ shots.

\subsection{Benchmarking LRE with GHZ-like circuits}\label{sec:ghz-circuits}

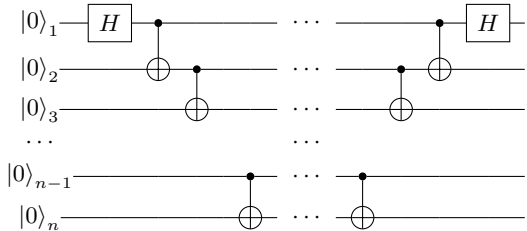
\begin{figure}[!htpb]
\begin{quantikz}[thin lines,
row sep=0.2cm,
column sep=0.2cm]
\ket{0}_1 & 
\gate{H} & 
\ctrl{1} &
\qw &
\qw &
\qw &
\qw &
\cdots & {}&
\qw &
\qw &
\ctrl{1} &
\gate{H} &
\qw \\
\ket{0}_2 & 
\qw & 
\targ{} & 
\ctrl{1} &
\qw &
\qw &
\qw &
\cdots & {}&
\qw &
\ctrl{1} &
\targ{} &
\qw &
\qw \\
\ket{0}_3 & 
\qw & 
\qw &
\targ{} &
\qw &
\qw &
\qw &
\cdots & {}&
\qw &
\targ{} &
\qw &
\qw &
\qw \\
\cdots&&&&&&&\cdots&\\
\ket{0}_{n-1} &
\qw & 
\qw &
\qw &
\qw &
\ctrl{1} &
\qw &
\cdots & {}&
\ctrl{1} &
\qw &
\qw &
\qw &
\qw \\
\ket{0}_n & 
\qw & 
\qw &
\qw &
\qw &
\targ{} &
\qw &
\cdots & {}&
\targ{} &
\qw &
\qw &
\qw &
\qw
\end{quantikz}
\caption{A GHZ-like benchmarking circuit composed of an $n$-qubit GHZ circuit followed by its inverse. By construction, the expectation value of $O = \ketbra{00\dots}{00\dots}$ evaluated on an ideal noiseless device is equal to 1.}
\label{fig:ghz_ghz_inverse_circ}
\end{figure}

The first type of benchmark circuit that we use to test LRE is based on the concatenation of a GHZ circuit followed by its inverse, as shown in Figure~\ref{fig:ghz_ghz_inverse_circ}.

The intermediate states during the execution of a GHZ-like circuit are highly entangled and, therefore, highly sensitive to environmental noise and decoherence. For this reason, they provide a good playground for testing the efficacy of LRE on structured, entangling circuits.

\subsubsection{Vary over number of layers}\label{sec:ghz-circuits-num-qubits}

\begin{figure}[!htpb]
    \centering
    \includegraphics[scale=0.5]{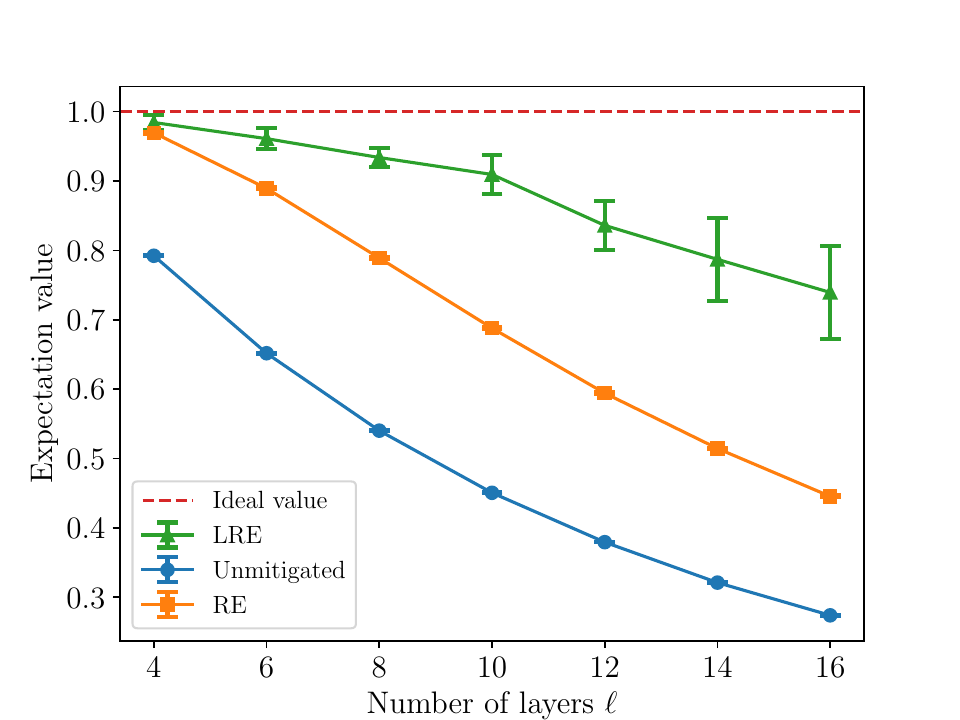}
    \caption{Expectation value of the observable $O = \ketbra{00\dots}{00\dots}$ estimated with different error mitigation strategies for a GHZ-like circuit as defined in Figure~\ref{fig:ghz_ghz_inverse_circ}. Each data point is averaged over 10 trials. For each trial, a total budget of $s_{\rm tot}=10^6$ shots is used. Error bars for each data point represent the standard deviation over the 10 independent trials. For all the data points considered in this example, layerwise Richardson extrapolation (LRE) is more accurate than traditional single-variable Richardson extrapolation (RE) and direct unmitigated estimation.}
    \label{fig:ghz-num-qubits}
\end{figure}

In Figure~\ref{fig:ghz-num-qubits} and Table~\ref{tab:ghz-num-qubits}, we compare the performance of LRE relative to RE and the unmitigated case as the number of layers $l$ increases (the number of qubits increases as well since $l=2n$). 

Increasing the size of a GHZ circuit elevates its complexity and susceptibility to errors. As expected, the estimation error increases with $l$ for all the results but, for each $l$, the expectation value estimated with LRE is closer to the ideal value. 
Error bars are evaluated by repeating the same experiment for 10 trials and computing the standard deviation of the results. We observe that LRE results are subject to higher statistical uncertainty. This is expected from the overhead analysis presented in Section \ref{sec:overhead} and from Figure~\ref{fig:overhead}: the error mitigation cost $c$ of LRE increases with the number of layers (or circuit chunks) and, for a fixed budget of shots $s_{\rm tot}=10^6$, this implies a proportional increase of the statistical variance. 
Note however that, even taking into account error bars,  the overall estimation error of LRE is smaller than RE due to a strong reduction of the estimation bias.

\setlength{\tabcolsep}{0.45em}
    \begin{table}[h!]
        \centering
        \begin{tabular}{c|c|c|c|c}
         Depth & Unmitigated & RE & LRE & Improvement \\ \hline 
    	 2 & 0.2078 & 0.0306 & 0.0174 & 75.41\% \\ 
    	 3 & 0.3483 & 0.1107 & 0.0390 & 183.75\% \\ 
    	 4 & 0.4599 & 0.2110 & 0.0662 & 218.79\% \\ 
    	 5 & 0.5495 & 0.3121 & 0.0906 & 244.34\% \\ 
    	 6 & 0.6206 & 0.4058 & 0.1640 & 147.40\% \\ 
    	 7 & 0.6789 & 0.4856 & 0.2130 & 127.98\% \\ 
    	 8 & 0.7261 & 0.5546 & 0.2607 & 112.76\%
    \end{tabular}
    \caption{Table of mean absolute estimation errors for each data point reported in Figure~\ref{fig:ghz-num-qubits}. The last column provides a percentage of improvement for the performance of layerwise Richardson extrapolation (LRE) over single-variable Richardson extrapolation (RE).  We observe that the performance improvement is significant even if the noise model (amplitude damping) is fixed and uniform along the circuit. This fact can be explained by noticing that, due to the specific structure of the circuit, noise has different impacts on the final expectation value depending on which specific layer it acts on. On the contrary, RE is completely insensitive to such fine-grained resolution of the noise impact.}
    \label{tab:ghz-num-qubits}    
\end{table}

\subsubsection{Vary over extrapolation order}\label{sec:ghz-circuits-degree}

\begin{figure}[!h]
    \centering
    \includegraphics[scale=0.5]{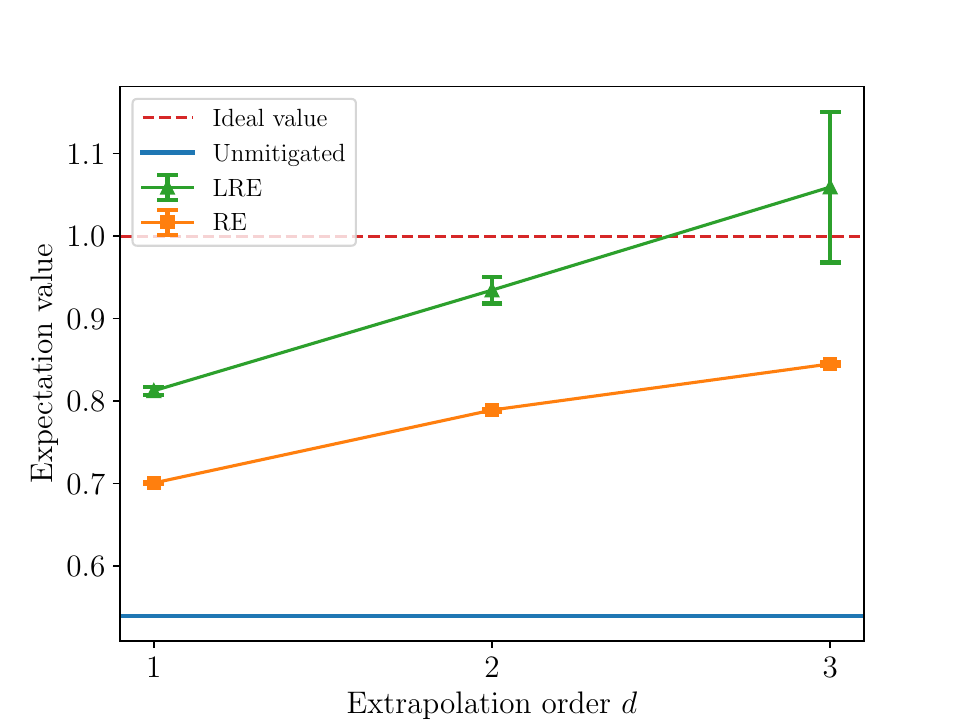}
    \caption{
    Expectation value estimated with layerwise Richardson extrapolation (LRE) and single-variable Richardson extrapolation (RE) for different values of the extrapolation degree $d$ (linear, quadratic, cubic). As a benchmark circuit we used a 4-qubit GHZ-like circuit having the structure shown in Figure~\ref{fig:ghz_ghz_inverse_circ} and as observable we used $O = \ketbra{00\dots}{00\dots}$. The error bars are obtained by calculating the standard deviation over $10$ trials. For each trial, a fixed budget of $s_{\rm tot}=10^6$ shots is used.}
    \label{fig:ghz-num-degree}
\end{figure}

In Figure~\ref{fig:ghz-num-degree}, we explore how the performance of LRE and RE varies with the extrapolation order $d$, i.e., the degree of the interpolating polynomial. Specifically, for both LRE and RE, we compare the results obtained via linear, quadratic, and cubic extrapolation.  As expected, the bias of both LRE and RE decreases with the extrapolation order $d$. However, statistical noise increases (exponentially) with $d$. In practice, for real-world use cases, we expect LRE and RE to be useful for $ 1 \le d \le 3$, since large values of $d$ are subject to the instabilities typical of high-order polynomial interpolation.

For applications where high fidelity is paramount, and resource constraints are less stringent, high extrapolation orders (e.g. quadratic or cubic) may be preferable. Conversely, for more resource-constrained environments or where moderate improvements in fidelity are sufficient, low extrapolation orders  (e.g. linear)  might be more suitable.

\subsubsection{Vary over number of shots}\label{sec:ghz-circuits-shots}

\begin{figure}[!b]
    \centering
    \includegraphics[scale=0.5]{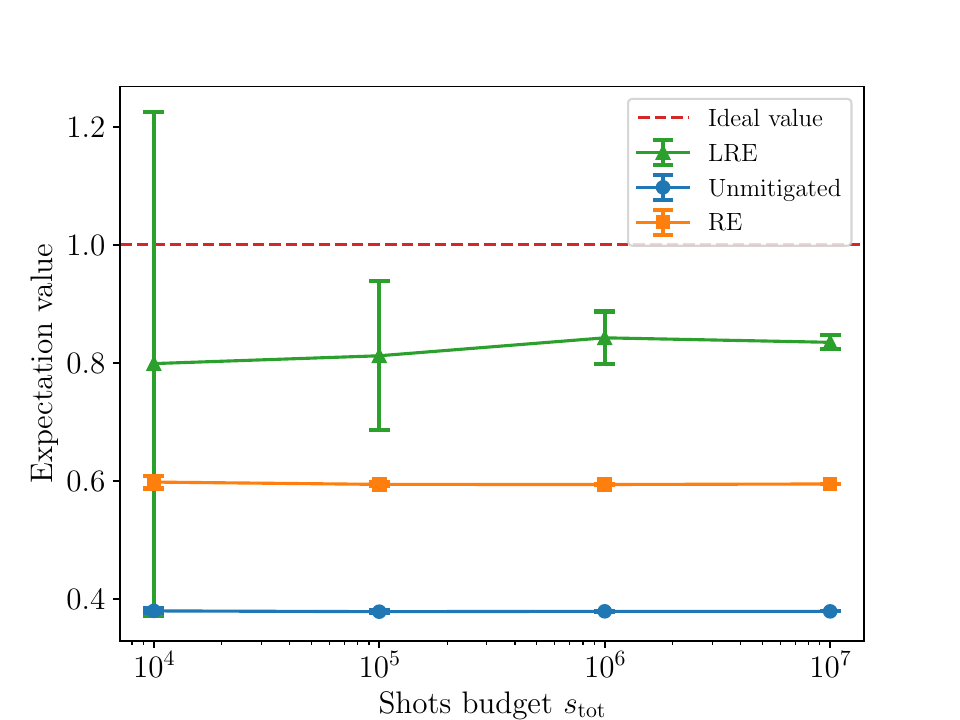}
    \caption{Expectation value of the observable $O = \ketbra{00\dots}{00\dots}$ estimated with different error mitigation strategies for a $6$-qubit GHZ-like circuit as defined in Figure~\ref{fig:ghz_ghz_inverse_circ}. Each data point is averaged over 10 trials. For each trial, we use a budget of shots as reported in the horizontal axis. The error bars in each data point illustrate the standard deviation over the 10 trials.}
    \label{fig:ghz-num-shots}
\end{figure}

In the previous simulations, for each expectation value estimation, we used a fixed budget of $s_{\rm tot}=10^6$ shots (total number of circuit executions). 
We now analyze what happens if we vary $s_{\rm tot}$. The results are depicted in Figure~\ref{fig:ghz-num-shots}.

Increasing the number of shots induces a reduction of the statistical variance for any estimation strategy (unmitigated, LRE, RE). However, we observe that LRE is much more sensitive to statistical noise and, as a consequence, to the number of shots.
For a small number of shots, the statistical variance of LRE is too large to produce a reliable estimation. In this regime, one could try to reduce the overhead of LRE by reducing the number of chunks $l$ or by increasing the gap $\Delta$ between scale factors.  As the number of shots increases, the performance of LRE stabilizes, yielding more consistent and reliable results.

\subsubsection{Vary over the gap between scale factors}\label{sec:ghz-circuits-scale-factor-multiplier}

\begin{figure}[!htpb]
    \centering
    \includegraphics[scale=0.5]{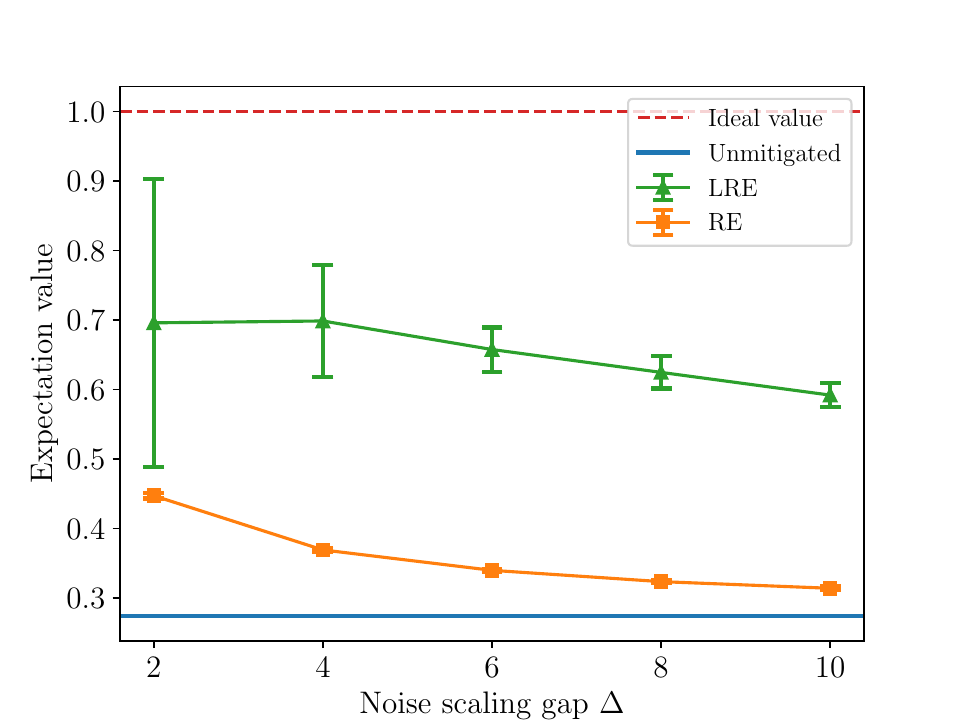}
    \caption{Expectation value of the observable $O = \ketbra{00\dots}{00\dots}$ estimated with increasing gap $\Delta$ between scale factors,  for an $8$-qubit GHZ-like circuit. We assume a fixed and limited number of shots $s_{\rm tot}=10^5$ for any estimation strategy. The error bars are obtained by calculating the standard deviation over the $10$ trials.}
    \label{fig:ghz-scale-factor-multiplier}
\end{figure}

This section delves into the impact of increasing the minimum gap $\Delta$ between noise scale factors, as a way of reducing statistical noise at the cost of increasing the estimation bias. 
The results are presented in Figure~\ref{fig:ghz-scale-factor-multiplier}.

From our previous analysis of the sampling overhead of LRE, we know that increasing the gap between noise scale factors reduces the sampling cost (see Figure~\ref{fig:overhead_vs_lambda_min}). Equivalently, for a fixed number of shots $s_{\rm tot}$, we expect a reduction of statistical noise for larger values of $\Delta$. This is indeed what we observe for LRE  in Figure~\ref{fig:ghz-scale-factor-multiplier}, where error bars get smaller for increasing $\Delta$. A similar reduction is also present for single-variable RE but, error bars are too small to be visible in the plot. In Figure~\ref{fig:ghz-scale-factor-multiplier} we also see the drawback of using a large gap between noise scale factors: the bias of the associated extrapolation increases due to stronger noise amplification. For practical scenarios, we expect that the net effect of increasing $\Delta$ is typically not a convenient strategy when using traditional RE, but it can help when using LRE due to its larger sampling cost.

\subsubsection{Vary over the number of circuit chunks }\label{sec:chunk-experiments}

Finally, we explore the influence of varying the number of circuit chunks $l$  on the estimation accuracy of LRE. 
As discussed in Section~\ref{sec:chunk-theory}, we are not forced to apply LRE to depth-1 layers, but we can apply it to multi-layer chunks of the input circuit. This implies that we are free to split the circuit into an arbitrary number $l=1, 2, \dots, l_{\rm max}$ of chunks, where the upper limit $l_{\rm max}$ is the total number of depth-1 layers.

\begin{figure}[!htpb]
    \centering
    \includegraphics[scale=0.5]{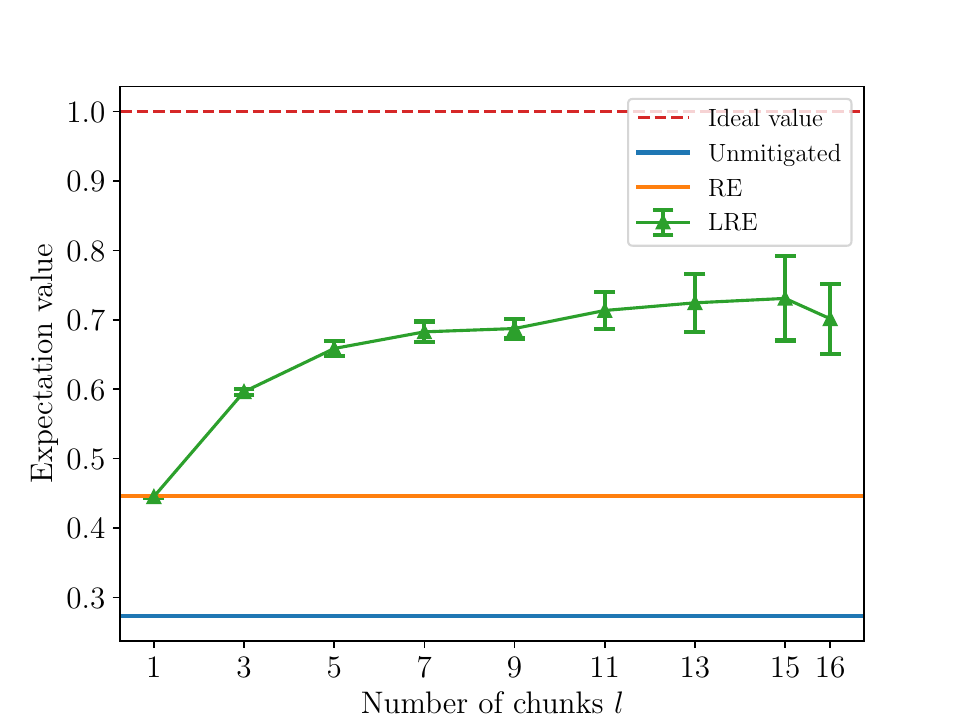}
    \caption{Expectation value of the observable $O = \ketbra{00\dots}{00\dots}$ evaluated via layerwise Richardson extrapolation (LRE) as a function of the number of chunks into which the input circuit is split (as proposed in Section~\ref{sec:chunk-theory}). As a benchmark circuit, we use an 8-qubit GHZ-like circuit. The blue line represents the unmitigated expectation value, the orange line depicts the result of applying single-variable RE, and the green triangles show the results after applying LRE. As expected, LRE reduces to RE for $l=1$. Error bars report the standard deviation over 10 trials. For each trial, a fixed budget of $s_{\rm tot}=10^6$ shots is used.}
    \label{fig:ghz-num-chunks}
\end{figure}

In Figure~\ref{fig:ghz-num-chunks}, we apply LRE to an 8-qubit GHZ-like circuit assuming a splitting of the circuit into a different number of chunks. In practice, for each $l$, we split the circuit into $l$ chunks of approximately equal depth (up to a rounding error of at most a single layer). Afterward, we apply LRE in the same way as in the previous examples but, instead of associating a noise scale factor to each depth-1 layer, we associate a scale factor to each chunk. For each circuit chunk, we use \emph{local folding} as defined in Equation~\eqref{eq:local_folding}  (also employed for RE). 

By construction, LRE reduces to RE for $l=1$. For larger values of $l$, we observe a significant reduction of the bias for LRE. We also observe an increase in statistical noise for large $l$, as expected. The overall interpretation of Figure~\ref{fig:ghz-num-chunks} is that, if we can afford the sampling overhead, it is always convenient to increase $l$. However, we also expect that for deeper circuits (e.g. $l_{\rm max}>100$) the complexity and the sampling cost of applying LRE at the level of single layers may become too large such that applying LRE on a smaller number of multi-layer chunks is a more pragmatic solution.

\subsection{Benchmarking LRE with randomized  circuits}\label{sec:random-cnot-circuit}

\begin{figure}[!htpb]
\begin{quantikz}[thin lines,
row sep=0.2cm,
column sep=0.2cm]
\ket{0}_1 & 
\gate{H} & 
\ctrl{1} &
\qw &
\qw &
\targ{} & 
\qw &
\qw \\
\ket{0}_2 & 
\ctrl{1} & 
\targ{} & 
\qw &
\ctrl{1} &
\qw &
\targ{} &
\qw \\
\ket{0}_3 & 
\targ{} & 
\gate{Y} &
\targ{} &
\targ{} &
\qw &
\ctrl{-1} &
\qw \\
\ket{0}_{4} &
\qw & 
\qw &
\ctrl{-1} &
\qw &
\ctrl{-3} &
\qw &
\qw \\
\end{quantikz}
\caption{An example of a randomly generated $4$-qubit circuit $C_{\rm rand}$, with high CNOT density. Note that the actual circuit used to benchmark LRE is $C=C_{\rm rand}^{-1} C_{\rm rand}$. }
\label{fig:random-circuit-example}
\end{figure}
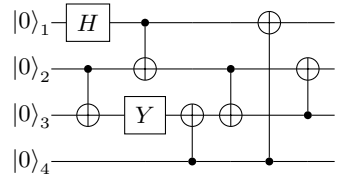

In this subsection, we use a different benchmark circuit to test the error mitigation performance of LRE. 
Instead of the GHZ-like circuit used in the previous examples, we use randomized circuits having the following structure:
\begin{equation}\label{eq:randomized-circuit}
C = C_{\rm rand}^{-1} C_{\rm rand},
\end{equation}
where $C_{\rm rand}$ is a random circuit obtained via a randomized application of single-qubit gates (${H, X, Y, Z, S, T}$) and CNOT gates. An instance of $C_{\rm rand}$ is shown in Figure~\ref{fig:random-circuit-example}.
To increase the amount of entanglement during the circuit execution, we assign a high probabilistic weighting to CNOT gates ($p_{\text{CNOT}} = 0.9$), thus ensuring a high density of CNOT gates in the benchmark circuits.

\begin{figure}[!b]
    \centering
    \includegraphics[scale=0.5]{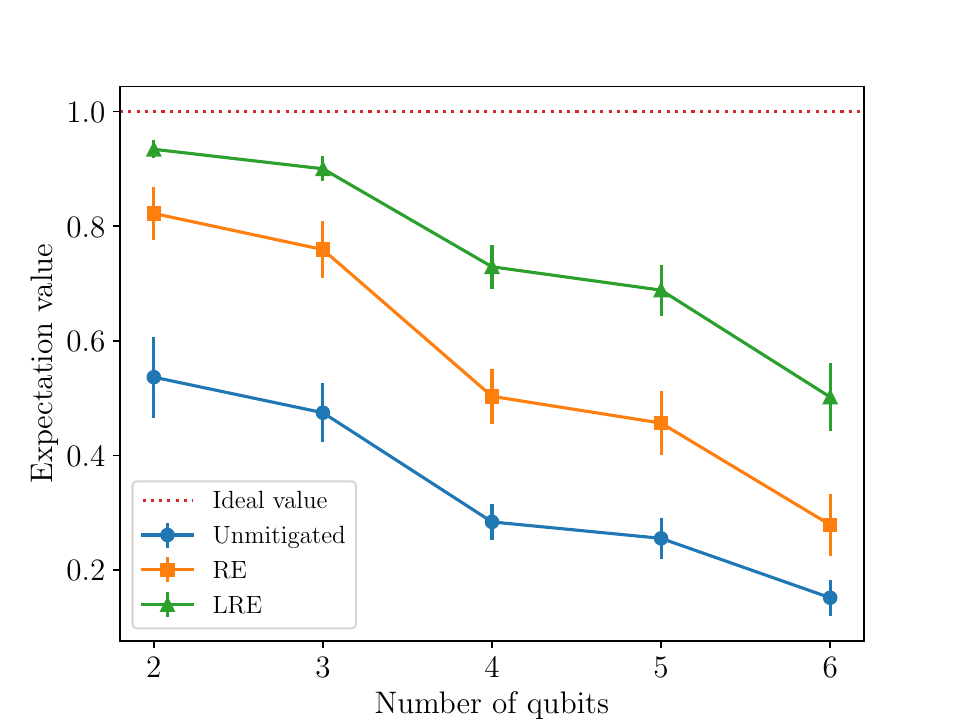}
    \caption{Expectation value of the observable $O = \ketbra{00\dots}{00\dots}$ estimated with different error mitigation strategies for a randomized circuit as defined in Equation~\eqref{eq:randomized-circuit} having total depth $l_{\rm max}=4$. Each data point is averaged over 10 different random instances of the benchmark circuit. Error bars for each data point represent the standard deviation over the 10 random instances.  For each circuit, a total budget of $s_{\rm tot}=10^6$ shots is used.  
    }
    \label{fig:LRE-random-circuit}
\end{figure}

Unlike GHZ-like circuits in which the depth is implied by the number of qubits, for the randomized circuits considered in this subsection, we are free to independently vary the number of qubits and the number of layers.
This freedom allows us to explore the performance of LRE when varying the number of qubits (at constant depth).

Figure~\ref{fig:LRE-random-circuit} presents a comparative analysis of the error mitigation performance when varying the number of qubits. The plot aggregates results obtained across $10$ randomly generated circuits.
The results are qualitatively similar to the GHZ-like case reported in Figure~\ref{fig:ghz-num-qubits}, in the sense that LRE outperforms both single-variable RE and the trivial unmitigated estimation. However, we also notice an important difference: error bars do not grow when increasing the number of qubits. This is a characteristic feature of Richardson extrapolation (both LRE and RE), for which the sampling overhead only depends on the choice of noise scale factors. This implies that the overhead of LRE depends only on the depth $l$ (or number of circuit chunks) but not on the number of qubits.
Even if the statistical variance is constant concerning the number of qubits, the bias of all estimation strategies (LRE, RE, unmitigated) gets larger for wider circuits. 

\section{Discussion}\label{sec:discussion}

We introduced \emph{layerwise Richardson extrapolation} (LRE), an error mitigation technique inspired by conventional (single-variable) Richardson extrapolation (RE)~\cite{temme2017error, kandala2019error, endo2018practical} but generalized to a framework in which the errors acting on different layers of a circuit can be amplified independently. We then presented several numerical experiments in which we compared LRE against conventional RE and direct unmitigated estimation.

Our findings suggest that LRE can be a convenient technique for practical applications since it presents several advantages (low bias, flexible sampling cost, noise-model agnostic). The main limitations of LRE are its statistical uncertainty (higher than RE) and the requirement of running a significant number of different circuits (similar to PEC \cite{temme2017error, endo2018practical}). We also explored different ways of reducing the sampling cost, such as increasing the gap between scale factors or reducing the number of circuit splittings, that can be useful for controlling the balance between error mitigation bias and sampling cost in large-scale experiments. 
From a theoretical perspective, LRE also provides a general multivariate formalism in which previous techniques are recovered as special limits. For example, LRE reduces to conventional RE for $l=1$ and to the noise-scaling version of the NOX protocol ~\cite{ferracin2022efficiently} for $d=1$.

The new technique proposed in this work opens up avenues for further research. 
In our examples, we considered numerical experiments based on a simple amplitude-damping noise model. It would be interesting to numerically investigate other noise models or, even better, test LRE on real hardware.
An aspect worth exploring is the design of suitable calibration experiments~\cite{calderon2023quantum, patel2022charter, hour2024improving, van2023probabilistic, ferracin2022efficiently} to estimate the noise levels of different layers or to determine the optimal hyperparameters of the LRE protocol, given a specific backend. For example, one could run calibration experiments to optimize the noise scaling gap $\Delta$, the number of circuit splittings $l$, and the extrapolation order $d$.

An interesting analysis would be an experimental comparison between LRE and PEC \cite{temme2017error, endo2018practical}. In theory, PEC can provide a more tailored error mitigation, since it is a noise-aware technique while LRE is noise-agnostic. In practice, however, it is not obvious what technique is more competitive in a real-world scenario \cite{russo2023testing}. PEC requires many noise characterization experiments \cite{van2023probabilistic, ferracin2022efficiently} that are known to be complex, costly, and subject to imperfections which have a strong impact on the quality of the final result. LRE is instead simpler and perhaps more robust to imperfections since, by construction, the executed circuits are generated according to a noise-agnostic and deterministic protocol.  

Inspired by the PEC protocol, a future direction worth exploring is the probabilistic implementation of LRE. Rather than executing all the $M$ circuits necessary for computing the sum in Equation~\eqref{eq:lre-linear-combo}, a Monte Carlo method employing importance sampling could be utilized. This approach would selectively and probabilistically evaluate only a subset of the terms in the full sum and could potentially extend the applicability of LRE to more layers $l$ and to higher orders $d$.

\section*{Code availability}

Software that implements the LRE method along with the code that is used to generate the data and plots in this work is available in~\cite{unitary2023mlzne}.

\section*{Acknowledgments}

VR acknowledges Nate T. Stemen and Nathan Shammah for insightful discussions as well as William J. Zeng for suggesting the idea of applying layerwise folding as a tool for error mitigation. This work was supported by the U.S. Department of Energy, Office of Science, Office of Advanced Scientific Computing Research, Accelerated Research in Quantum Computing under Award Numbers DE-SC0020266 and DE-SC0020316 as well as by IBM under Sponsored Research Agreement No. W1975810. 

\twocolumngrid

\bibliographystyle{ieeetr}
\bibliography{refs.bib}

\onecolumngrid
\newpage 

\section{Appendix}

\subsection{Noise model for experiments}\label{sec:noise-model}

For the experiments in Section~\ref{sec:experiments}, we consider a noise model characterized by amplitude damping errors. Let the probability of amplitude damping error for a single qubit and two-qubit gate be denoted as $p_1$ and $p_2$ respectively, with $p_1 = 0.04$ and $p_2 = 0.08$. The single-qubit amplitude damping channel is represented as
\begin{equation}
    \mathcal{E}_1(\rho) = E_0 \rho E_0^\dagger + E_1 \rho E_1^\dagger
\end{equation}
where,
\begin{equation}\label{eq:kraus-appendix}
    E_0 = 
    \begin{bmatrix} 
    1 & 0 \\ 0 & \sqrt{1 - p_1} 
    \end{bmatrix} 
    \quad \text{and} \quad
    E_1 = 
    \begin{bmatrix} 
    0 & \sqrt{p_1} \\ 0 & 0 
    \end{bmatrix}.
\end{equation}
This channel is added to all single-qubit gates. For the two-qubit CNOT gate, we apply the tensor product of two single-qubit channels,
\begin{equation}
    \mathcal{E}_2(\rho) = \sum_{i \in \{0, 1\}} \sum_{j \in \{0, 1\}}E_i \otimes E_j \, \rho \, E_i^\dagger \otimes E_j^\dagger,   
\end{equation}
where we replace $p_1$ in Equation~\eqref{eq:kraus-appendix} with $p_2=0.08$ .
We use the Qiskit Aer simulator~\cite{qiskit} to simulate circuits with the above noise model.

\subsection{Multivariate Lagrange interpolation in LRE}\label{sec:multivariate-interpolation}

Single-variable  Lagrange interpolation constructs a single-variable polynomial to fit a set of $N$ points in $\mathbb{R}^2$~\cite{gasca2000polynomial}. In the case of \emph{multivariate Lagrange interpolation}, this approach is extended to handle the multivariate polynomial interpolation of points in higher-dimensional spaces. Here, for a set of $N$ points of a polynomial with $l$ variables, the interpolation is conducted in $\mathbb{R}^{l+1}$~\cite{saniee2008simple, olver2006multivariate}. In this appendix we adapt the mathematical formalism of Lagrange interpolation of \cite{saniee2008simple} to the specific notation of the LRE framework introduced in Section~\ref{sec:lre-extrapolation}. 

We aim to find the interpolating $l$-variable polynomial passing through a set of $N$ points representing the noise-scaled expectation values of an observable. Each of these points corresponds to a circuit execution under a specific noise scaling, captured by a vector $\boldsymbol{\lambda}$ containing $l$ real scale factors, corresponding to the amount of noise scaling applied to the $l$-th layer of the circuit. Given the $N$ measured points, we define the set of scale factor vectors as $\Lambda= \{ \boldsymbol{\lambda}_1, \boldsymbol{\lambda}_2, \dots, \boldsymbol{\lambda}_N \}$ and the array of the associated expectation values as
\begin{equation}\label{eq:exp-value-vector-appendix}
    \mathbf{z} = 
    \left(
        \langle O(\boldsymbol{\lambda}_1) \rangle,
        \langle  O(\boldsymbol{\lambda}_2) \rangle,
        \ldots,
        \langle O(\boldsymbol{\lambda}_N) \rangle
    \right)^{\t}.
\end{equation}
The most general $l$-variable polynomial of degree $d$ can be written as
\begin{equation}\label{eq:interpolating-polynomial-appendix}
    P(\boldsymbol{\lambda}) = \sum_{j=1}^M c_j \mathcal{M}_j(\boldsymbol{\lambda}, d),
\end{equation}
where $\{c_j\}$ are real coefficients and $\{ \mathcal{M}_j(\boldsymbol{\lambda}, d): j=1, 2, \dots, M\}$ is the set of all $l$-variable monomials of degree at most $d$. The number of monomials is given by $M=\binom{d+l}{d}$ and is therefore fixed by $l$ and $d$.
The interpolation problem corresponds to determining the $M$ unknown coefficients $\{ c_j\}$ such that the polynomial passes through the measured points, i.e.:
\begin{equation}
    P(\boldsymbol{\lambda}_i) =\langle O(\boldsymbol{\lambda}_i) \rangle = \mathbf{z}_i , \quad \forall \boldsymbol{\lambda}_i \in \Lambda.
\end{equation}
Define the following \emph{sample matrix} which contains the values of all monomials evaluated at each scale factor vector in $\Lambda$:
\begin{equation}\label{eq:sample-matrix-appendix}
    \mathbf{A}(\Lambda, d) = 
    \begin{bmatrix}
         \mathcal{M}_1(\boldsymbol{\lambda}_1, d) & \mathcal{M}_2(\boldsymbol{\lambda}_1, d)& \cdots & \mathcal{M}_M(\boldsymbol{\lambda}_1, d) \\
        \mathcal{M}_1(\boldsymbol{\lambda}_2, d) & \mathcal{M}_2(\boldsymbol{\lambda}_2, d)& \cdots & \mathcal{M}_M(\boldsymbol{\lambda}_2, d) \\
        \vdots & \vdots & \ddots & \vdots \\
        \mathcal{M}_1(\boldsymbol{\lambda}_N, d) & \mathcal{M}_2(\boldsymbol{\lambda}_N, d)& \cdots & \mathcal{M}_M(\boldsymbol{\lambda}_N, d) \\
    \end{bmatrix}.
\end{equation}
If we cast the coefficients of the polynomial in a vector $\mathbf{c}=(c_1, c_2, \dots, c_M)^\t$, the interpolation problem can be expressed as the following linear system:
\begin{equation}
 \mathbf{A} \mathbf{c} = \mathbf{z}.
\end{equation}
To have a unique solution, we assume $N=M$ and $\det(\mathbf{A}) \neq 0$.
In practice, given $l$ and $d$, this is a constraint on the number and the values of the scale factor vectors in the set $\Lambda$ that is straightforward to check and satisfy.

One way of determining the interpolating polynomial would be to solve for $\mathbf{c}$ and to replace the solution into Equation~\eqref{eq:interpolating-polynomial-appendix}. There is however an alternative way, which does not require the explicit computation of $\mathbf{c}$ and is given by the following Lagrange interpolation formula \cite{saniee2008simple}:

\begin{equation}\label{eq:lagrange-appendix}
P(\boldsymbol{\lambda}) = \sum_{i=1}^M \langle O (\boldsymbol{\lambda}_i) \rangle \frac{\det \left(\mathbf{M}_i (\boldsymbol{\lambda}) \right)}{\det \left(\mathbf{A}\right)} ,
\end{equation}
where $\mathbf{M}_i (\boldsymbol{\lambda})$ is the matrix obtained by substituting the $i$-th row of the sample matrix $\mathbf{A}$ with the same row of monomials but evaluated on the generic polynomial variable $\boldsymbol{\lambda}$ (instead of $\boldsymbol{\lambda}_i \in \Lambda$), for example:
\begin{equation}\label{eq:monomial-matrix-appendix}
    \mathbf{M}_2(\boldsymbol{\lambda}) = 
    \begin{bmatrix}
         \mathcal{M}_1(\boldsymbol{\lambda}_1, d) & \mathcal{M}_2(\boldsymbol{\lambda}_1, d)& \cdots & \mathcal{M}_M(\boldsymbol{\lambda}_1, d) \\
        \mathcal{M}_1(\boldsymbol{\lambda}, d) & \mathcal{M}_2(\boldsymbol{\lambda}, d)& \cdots & \mathcal{M}_M(\boldsymbol{\lambda}, d) \ \\
        \vdots & \vdots & \ddots & \vdots \\
        \mathcal{M}_1(\boldsymbol{\lambda}_N, d) & \mathcal{M}_2(\boldsymbol{\lambda}_N, d)& \cdots & \mathcal{M}_M(\boldsymbol{\lambda}_N, d) \\
    \end{bmatrix}.
\end{equation}
By construction, the right-hand side of Equation~\eqref{eq:lagrange-appendix} is a polynomial in the variable $\boldsymbol{\lambda}$ of degree at most $d$. Moreover, it is easy to check that it also interpolates all points since, if we evaluate the expression at a specific $\boldsymbol{\lambda}_j \in \Lambda$, we have 
\begin{equation}
    P(\boldsymbol{\lambda}_j) =  \sum_{i=1}^M \langle O (\boldsymbol{\lambda}_i) \rangle \frac{\det \left(\mathbf{M}_i (\boldsymbol{\lambda}_j) \right)}{\det \left(\mathbf{A}\right)} 
    =\langle O (\boldsymbol{\lambda}_j) \rangle \frac{\det \left(\mathbf{M}_j (\boldsymbol{\lambda}_j) \right)}{\det \left(\mathbf{A}\right)} 
    =\langle O (\boldsymbol{\lambda}_j) \rangle \frac{\det \left(\mathbf{A}\right) }{\det \left(\mathbf{A}\right)}= \langle O(\boldsymbol{\lambda}_j) \rangle,
\end{equation}
where we used that, for $i\neq j$, $\det \left(\mathbf{M}_i (\boldsymbol{\lambda}_j)\right)=0$ since the $i$-th row and the $j$-th row are equal.

Evaluating Equation~\eqref{eq:lagrange-appendix} at the zero-noise limit (denoted as $\boldsymbol{\lambda} = \mathbf{0}$), we get:
\begin{equation}\label{eq:lre-observable-appendix}
    O_{\rm LRE} = P(\mathbf{0}) = \sum_{i=1}^M \langle O (\boldsymbol{\lambda}_i)\rangle  \frac{\det \left(\mathbf{M}_i (\boldsymbol{0}) \right)}{\det \left(\mathbf{A}\right)}.
\end{equation}
The matrix $\mathbf{M}_i(\mathbf{0})$ can be obtained from the sample matrix $\mathbf{A}$ after replacing the $i$-th row by the array $\mathbf{e}_1=(1, 0, \ldots, 0)^\t$, since all monomials are zero at $\boldsymbol{\lambda}=\mathbf{0}$, with the exception of the constant one   $\mathcal M_1(\mathbf{0}, d)=\mathcal M_1(\boldsymbol{\lambda}, d)=1$. Here, we implicitly assumed that monomials are ordered with increasing degree. Otherwise, the element $1$ in the vector $\mathbf{e}_1$ should be shifted to the position associated with the zero-order monomial. Equation~\eqref{eq:lre-observable-appendix} corresponds to Equations~\eqref{eq:lre-linear-combo} and \eqref{eq:eta-coefficients} of the main text.

\end{document}